\documentclass[12pt]{article}
\usepackage{epsfig}  
\def\bea{\begin{eqnarray}} \def\eea{\end{eqnarray}}
\def\w{\u{u} }  
\begin{document}
\title{\bf Movoj sub newtona logaritma potencialo \\}  
\author{ A.F.F. Teixeira \thanks{teixeira@cbpf.br} \\
         {\small Centro Brasileiro de Pesquisas F\'{\i}sicas } \\ 
         {\small 22290-180 Rio de Janeiro-RJ, Brazil} } 
\date{16$^a$ de Oktobro de 2003}
\maketitle 

\abstract{A linear density of mass is assumed uniformly distributed along an infinitely long straight line. 
All possible motions of a test particle are described, under the newtonian logarithmic gravitational potential alone. 
Those most interesting occur in planes normal to the line. 
An infinity of closed orbits are found, and classified in terms of two positive integers. 

-----

Maso kun unuforma linia denseco estas ordata sur nefinita rekto. 
\^Gi produktas newtona logaritma potencialo.  
\^Ciu ebla movo de test-parteto estas studata. 
La plej interesaj movoj okazas en plano normala al linio de maso.  
Nefinita qvanto de fermaj trajektorioj estas findataj, kaj klasifikataj per du pozitivaj entjeroj. } 
\section{Anta\u{u}parolo}
Sendube la plej studata newtona potencialo en fiziko estas tiu de punkta parteto kun maso $m$.    
La hiperbola potencialo estas $-Gm/r$, kun $G$ universala konstanto kaj $r$ la distanco al maso.   
\^Ci tie potencialo havas sferan simetrion, kaj permesis kalkuli movojn de sunaj planedoj kun mirinda precizeco. 
 
Anka\w la plana kaj la cilindra estas gravaj spacaj simetrioj.  
La plana newtona potencialo estas donata de surfaca denseco de maso $\sigma$, unuforme ordata sur nefinita plano.  
La potencialo estas $2G\sigma h$, kun $h$ la distanco al la plano.   
 
La alia potencialo, logaritma kaj kun cilindra simetrio, estas la celo de jena artikolo.  
\^Gi korespondas al nefinita rekto kun unuforma linia denseco de maso $\lambda$.   
Por kompreni detalojn de \^gia gravito, ni studas la movon de test-partetoj.   
 
Ni unue studas la plej simplan movon, la solan radian. 
Sekvante ni studos movon sur iu plano kiu enhavas la linian mason. 
Poste ni vidos movojn sur plano normala al maslinio -- tiuj estas la plej interesaj.   
Fine, ni vidos pli \^generalan movon, sur rektata surfaco paralela al maslinio.  
\section{Logaritma potencialo} 
Estu nefinita linio de maso kun denseco $\lambda$, sur akso $z$ de cilindra sistemo de koordinato.   
La newtona gravita potencialo de $\lambda$ estas logaritma,  
\bea                                                                    \label{e01}
N(\rho)=2G\lambda\ln (\rho/{\rm const}), 
\eea
kun $G\approx 2/3\times 10^{-10}$ MKS la newtona gravito konstanto, kaj $\rho$ la radia cilindra koordinato.  
La potencialo $N(\rho)$ estas pozitive nefinita en radia infinito $\rho\rightarrow\infty$, kaj negative nefinita \^cirka\w la akso $\rho=0$. 
\section{Radia movo} 
Ni komencu studi movon sur linio normala al akso $z$. 
En $t=0$, estu senmova test-parteto en distanco $\rho\!=\!\rho_0$ de akso $z$.  
\^Gi movos la\w la direkto al akso kun akcelo  
\bea                                                                    \label{e02}
g(\rho)=-2G\lambda/\rho.  
\eea
Akcelo kreskas dum la falo, e\^c nefiniti\^gas \^cirka\w $\rho=0$. 
 
Dum la falo la tuta energio (kinetika plus potenciala) konservi\^gas,  
\bea                                                                    \label{e03} 
\frac{1}{2}\dot{\rho}^2 + 2G\lambda\ln (\rho/{\rm const})=0+2G\lambda \ln (\rho_0/{\rm const}); 
\eea
\^ci tie, punkto signifas tempa derivo. Do la parteta velo en $\rho$ estas  
\bea                                                                    \label{e04} 
\dot{\rho}=-\sqrt{4G\lambda\ln{\rho_0/\rho}}. 
\eea 
Vidu ke anka\w la velo estas nefinita \^cirka\w la akso $\rho=0$. 
Ni difinas $y:=\sqrt{\ln{\rho_0/\rho}}$ kaj solvas (\ref{e04}) per la eraro funkcio  
\bea                                                                    \label{e05} 
{\rm erf}(x):=\frac{2}{\sqrt{\pi}}\int_{0}^{x}{e^{-y^2}dy}. 
\eea 
Funkcio erf$(x)$ kreskas unuforme de erf(0)=0 \^gis erf($\infty)$=1. 
La solvo estas    
\bea                                                                    \label{e06}
t(\rho)=t_{\!f} \, {\rm erf}\left(\sqrt{\ln{\rho_0/\rho}}\right), \hskip3mm t_{\!f}:=\frac{\rho_0}{2}\,\sqrt{\frac{\pi}{G\lambda}}, 
\eea 
kaj estas videbla en bildo 1. Rimarku en (\ref{e06}) ke da\u{u}ro $t_{\!f}$ de falo kreskas linie kun komenca distanco $\rho_0$, kaj \^gia qvadrato malkreskas linie kun denseco de maso $\lambda$.

\vspace*{3mm}
\centerline{\epsfig{file=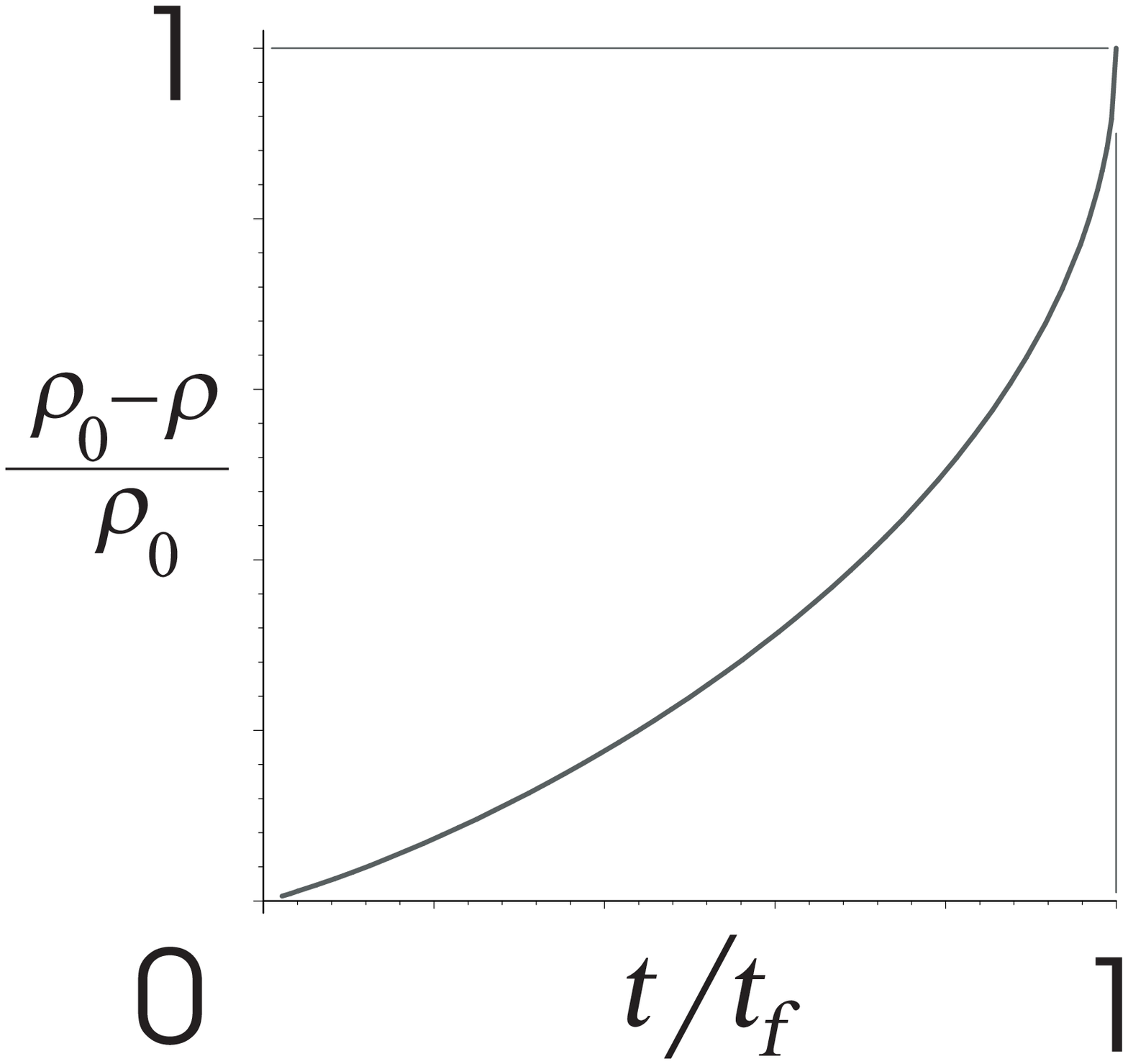,width=3cm,height=3cm}} 
\vspace*{3mm} 

\noindent {\small {\bf Bildo 1} {\it En $t=0$ estas senmova test-parteto je distanco $\rho=\rho_0$ de akso $z$. Gravito akcelas parteton kun intenseco pli kaj pli forta. Amba\w akcelo kaj velo i\^gas nefinitajn en $\rho\!=\!0,\, t\!=\!t_f$.  }}  
\vspace*{5mm}

Atentu ke (\ref{e06}) validas nur dum la falo, $0\leq t\leq t_f$. 
Se parteto ne halti\^gas en akso, \^gi iras rekte sed kun negativa akcelo, sekvante denove atingas distancon $\rho=\rho_0$ en $t=2t_f$, kaj komencas revenon al akso.  
Kaj cetere, en \^ciama movo kun periodo $4t_f$. 
\section{Movo en plano $\phi=$ konst}
La $z$-simetrio de gravita fonto faras konstantan la la\u{u}longa $v_z$. 
Movo de parteto en plano $\phi=$ konst do estas simpla kunmeto de radia movo (\ref{e06}) kaj rekta unuforma la\u{u}longa movo $z=v_zt+$konst. 

Se en $t=0$ parteto ekveturas de $\rho=\rho_0, z=0$, kun pure la\u{u}longa velo $\dot{z}=v_z$,  
tiuokaze la formo de \^gia trajektorio (vidu bildon 2) estas donata per  
\bea                                                                     \label{e07}
z(\rho)=v_z\,t_{\!f}\,{\rm erf}\left(\sqrt{\ln{\rho_0/\rho}}\right), \hskip3mm t_{\!f}:=\frac{\rho_0}{2}\,\sqrt{\frac{\pi}{G\lambda}}.  
\eea

\vspace*{3mm}
\centerline{\epsfig{file=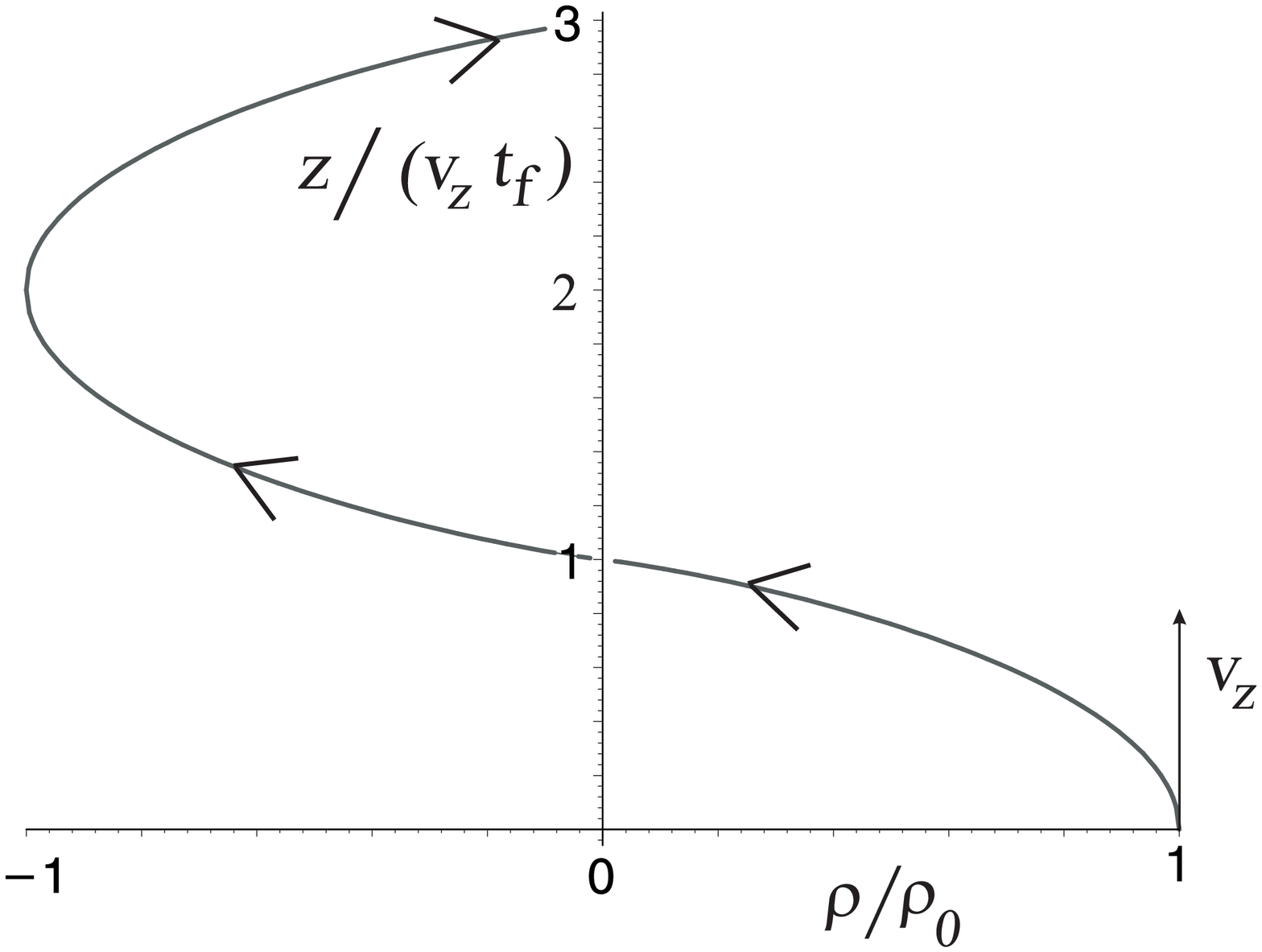,width=7cm,height=5cm}} 
\vspace*{3mm} 

\noindent {\small {\bf Bildo 2} {\it Trajektorio de parteto komencante de $\rho=\rho_0$, $z=0$ en $t=0$, kun pure la\u{u}longa velo $v_z$. 
La parteto atingas akson $z$ en $t=t_f$, perpendikle, sed la precizeco de la bildo ne permesas vidi tion.} } 
\vspace*{5mm} 

Se parteto ne halti\^gis en akso, \^gia movo estos serio de svingoj, la\u{u}longe akso $z$, ripetante kun interspacoj $\Delta z=4v_zt_f$. 

Ni scias ke la lumo veturas, en vakuo kaj sen gravito, kun konstanta velo $c$. 
Tiu estas baza fakto en speciala teorio de relativeco. 
Je la \^generala teorio de relativeco, kiu konsideras spacojn kun gravito, la velo de lumo anka\u{u} estas $c$, kondi\^ce ke \^gi estos mezurata per iloj en libera falo sub gravito.  

Anta\u{u}diroj de newtona teorio de gravito pri lumaj trajektorioj estas tre malsimilaj al de \^generala relativeco.  
Malsameco kreskas kun intenseco de gravita kampo.  
En jena problemo, la kampo estas tre forta cirka\w la akso $z$, do anta\u{u}diroj de la du teorioj povas esti tre malsimilaj.  
Fakte, lumo kiu havus trajektorion kiel en bildo 2, kun $v_z=c$, interplektus akson $z$ je intertempoj $2t_f$ la\w la newtona teorio. 
Anstata\u{u}e, la \^generala relativeco diras ke tio intertempo estas malpli granda, $t_f\sqrt{2}$, por problemoj kun malgranda masdenseco, $G\lambda/c^2<<1$. 
Oni dirus, bazante en \^ci tiu ekzemplo, ke einsteina gravito agas sur lumo pli forte ol newtona gravito.  
\section{Movo en plano $z=$ konst} 
Unue pensu pri grava speciala okazo, de {\bf ronda movo} de parteto \^cirka\w akso $z$.  
En \^ci tie movo, gravita forto $2Gm\lambda/\rho$ egalas centrifugan forton $m{v_c}^2/\rho$, do tangenta velo estas  
\bea                                                                 \label{e08}
v_c:=\sqrt{2G\lambda}\hskip1mm . 
\eea
Ni vidas ke \^ciu valoro por radio de rondo estas ebla.  

Ni vidu nun \^generalan okazon.  
Supozu ke en $t=0$ parteto estas en $\rho=\rho_0,\,\,\phi=0$, kun pure azimuta velo $v_0$, kaj kun $\dot{\phi}>0$. 
\^Ciu movo havas tri ne-nulaj parametroj $\lambda, \rho_0, v_0$.   
Konservo de energio kaj angula momento donas, respektive,   
\bea                                                               \label{e09} \frac{1}{2}\dot{\rho}^2 +  \frac{1}{2}(\rho\dot{\phi})^2 + 2G\lambda \ln(\rho/{\rm const})  = 0 + 
\frac{1}{2}{v_0}^2 + 2G\lambda \ln(\rho_0/{\rm const}) , 
\eea 
\vskip-8mm
\bea                                                                 \label{e10}
\rho(\rho\dot{\phi}) = \rho_0 v_0. 
\eea 

Formo de trajektorioj venos el (\ref{e08}) -- (\ref{e10})
\bea                                                                  \label{e11}
{\rm d}\phi =\pm \frac{\rho_0 {\rm d}\rho}{\rho^2\sqrt{(v_c/v_0)^2\ln(\rho_0/\rho)^2-({\rho_0}^2/\rho^2-1)}}.
\eea 
Difinu $x:=\rho/\rho_0$ kaj skribu  
\bea                                                                  \label{e12}
{\rm d}\phi =\pm \frac{{\rm d}x}{x^2\sqrt{1-1/x^2-(1/\nu)^2\ln x^2}},\hskip2mm \nu:=v_0/v_c;
\eea
en  d$\phi$, nur la vela parametro $\nu$ \^ceestas. 

Ni unue studos okazojn kun $\nu<1$; \^gi havas komencan tangentan velon $v_0$ malgranda ol la  $v_c$ de rondaj movoj.   
Kun gravita forto pli granda ol centrifuga forto, $\rho$ malkreskas de $\rho_0=:\rho_{max}$ \^gis minimumo $\rho_{min}>0$.  
Velo de parteto unuforme kreskas de $v_0=:v_{min}<v_c$ \^gis maksimumo $v_{max}>v_c$. 

Anta\w integri (\ref{e12}), kelkaj paroloj pri niaj parametroj estos utilaj.  
Kiam $\nu<1$, la tri parametroj uzotaj estas $\lambda, \rho_{max}, v_{min}$. 
La\w (\ref{e09}) kaj (\ref{e10}), ni havas  
\bea                                                                  \label{e13}
{v_{max}}^2-{v_{min}}^2=4G\lambda\ln(\rho_{max}/\rho_{min}), \hskip5mm 
\rho_{min}v_{max}=\rho_{max}v_{min}. 
\eea 
Malsame al sfera okazo, ni ne sukcesis atingi simplajn esprimojn por $\rho_{min}$ kaj $v_{max}$. 
Do ni eliminis $v_{max}$ en (\ref{e13}) kaj atingis $\rho_{min}$ kiel entenita funkcio de $\rho_{max}$ kaj $\nu$,  
\bea                                                                  \label{e14}      
\nu=\sqrt{\frac{\ln(\rho_{max}/\rho_{min})^2}{(\rho_{max}/\rho_{min})^2-1}} \hskip1cm (\nu<1).
\eea 
Grafika\^j\o de $\rho_{max}/\rho_{min}$ kiel funkcio de $\nu<1$ estas en bildo 3, kiu anka\w montras $\rho_{min}/\rho_{max}$ kiel funkcio de $\nu>1$. 
Konante $\rho_{max}/\rho_{min}$, havi $v_{max}$ per la dua (\ref{e13}) estas simpla. 
Indas diri, eqvacio (\ref{e14}) ne estas vera por $\nu>1$; fakte, la dua (\ref{e13}) donas  $\nu_{>1}=(\rho_{max}/\rho_{min})\nu_{<1}$. 

\vspace*{3mm}
\centerline{\epsfig{file=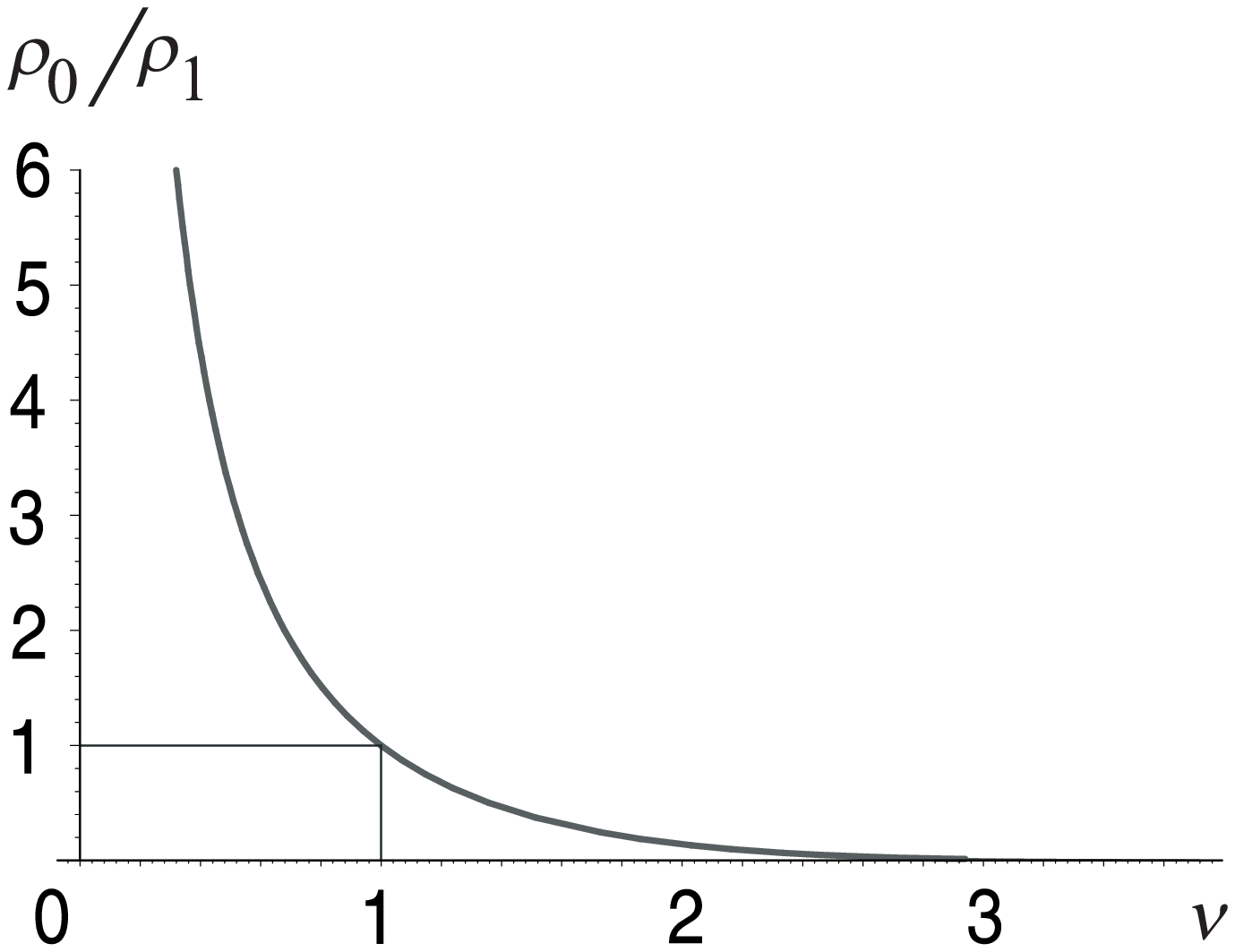,width=6cm,height=4cm}} 
\vspace*{3mm} 

\noindent {\small {\bf Bildo 3} {\it La parametro de velo $\nu=v_0/\sqrt{2G\lambda}$ estas 0-dimensia, kaj reprezentas la pure azimutan komencan velon $v_0$ de parteto. 
Kiam $\nu<1$, uzu $\rho_0=\rho_{max}$ kaj $\rho_1=\rho_{min}$. Kiam $\nu=1$, uzu   $\rho_0=\rho_1$, arbitraj, kaj la movo estas ronda. Kaj kiam $\nu>1$, uzu $\rho_0=\rho_{min}$ kaj $\rho_1=\rho_{max}$. } } 
\vspace*{5mm}
 
Nur ni revenu al integrado de (\ref{e12}), kun $\nu<1$. 
En komenca spaco inter $\rho=\rho_{max}$ kaj $\rho=\rho_{min}$, formo de trajektorio venas el  
\bea                                                                  \label{e15}
\phi(\rho)=\int_{\rho/\rho_{max}}^{1}{\frac{{\rm d}x}{x^2\sqrt{1-1/x^2-(1/\nu)^2\ln x^2}}}. 
\eea 
Ni ne atingis rigora integro de (\ref{e15}), do ni prezentas nur numeran solvon por fiksaj  valoroj de $\nu$. 
Malsame sub sfere simetrikaj potencialoj, nur escepte jenaj trajektorioj estos fermaj.  
Bildo 4 montras ekzemplon de ferman trajektorion. 

\vspace*{3mm}
\centerline{\epsfig{file=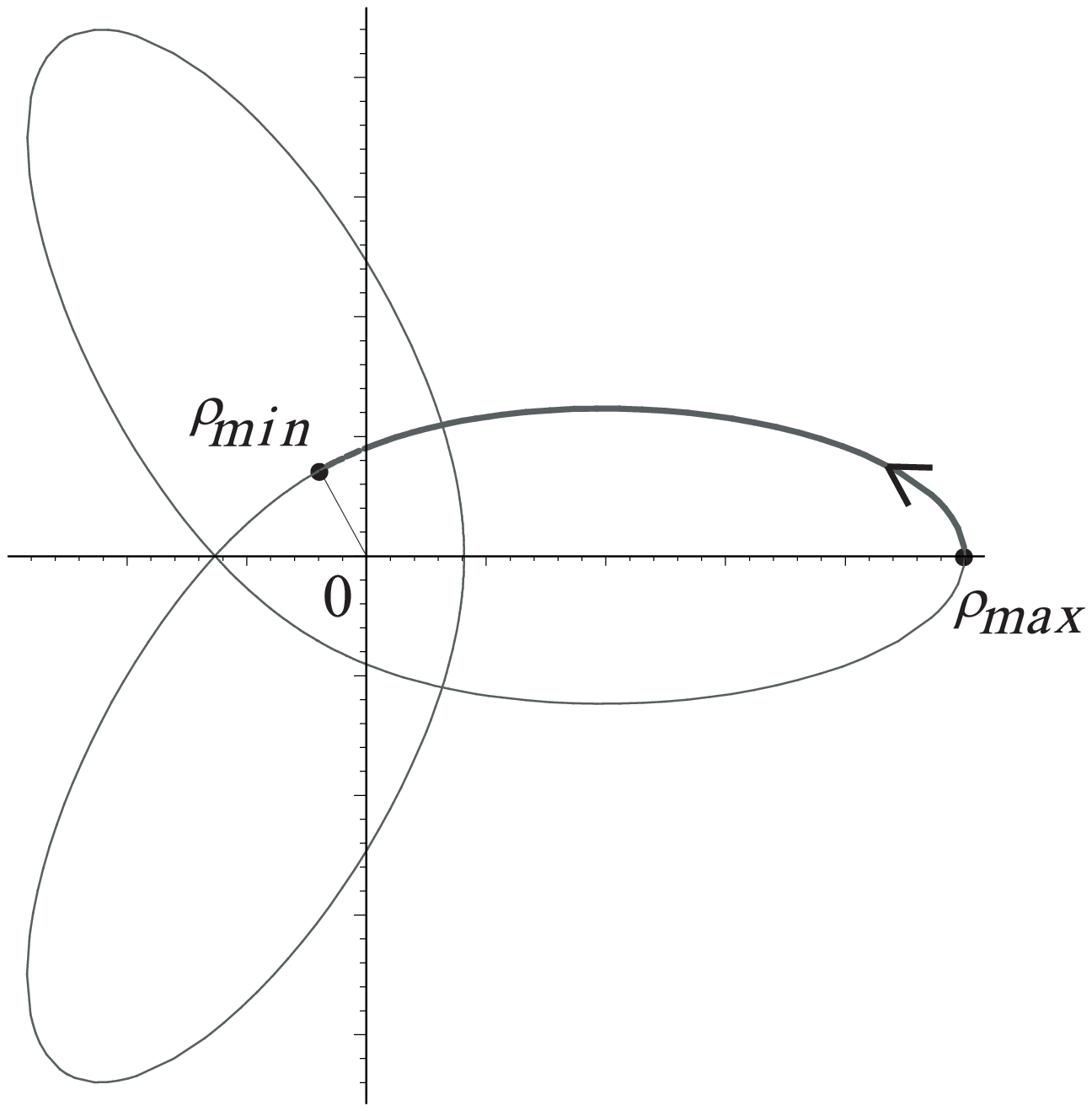,width=61mm,height=60mm}} 
\vspace*{3mm} 

\noindent {\small {\bf Bildo 4} {\it Ferma trajektorio de parteto kiu movas en plano $z=$ konst, kun $\nu\approx0,32$ (velo $v_{min}\approx0,32\sqrt{2G\lambda}$, en $\rho=\rho_{max}$). } }
\vspace*{5mm}

Angulo $\Phi$, mizurato inter duopo de sinsekvaj $\rho_{max}$ kaj $\rho_{min}$, estas klariganta  qvantumo.  
Je bildo 4 \^gi valoras $120^o$. La angulo $\Phi$ venas el (\ref{e15}) kiel   
\bea                                                                  \label{e16}
\Phi:=\int_{\rho_{min}/\rho_{max}}^{1}{\frac{{\rm d}x}{x^2\sqrt{1-1/x^2-(1/\nu)^2\ln x^2}}}. 
\eea
Racio $\rho_{min}/\rho_{max}$ estas funkcio de $\nu$, do ni povas grafiki $\Phi$ kiel funkcio de $\nu$ nur, vidu bildo 5. 
  
\vspace*{3mm}
\centerline{\epsfig{file=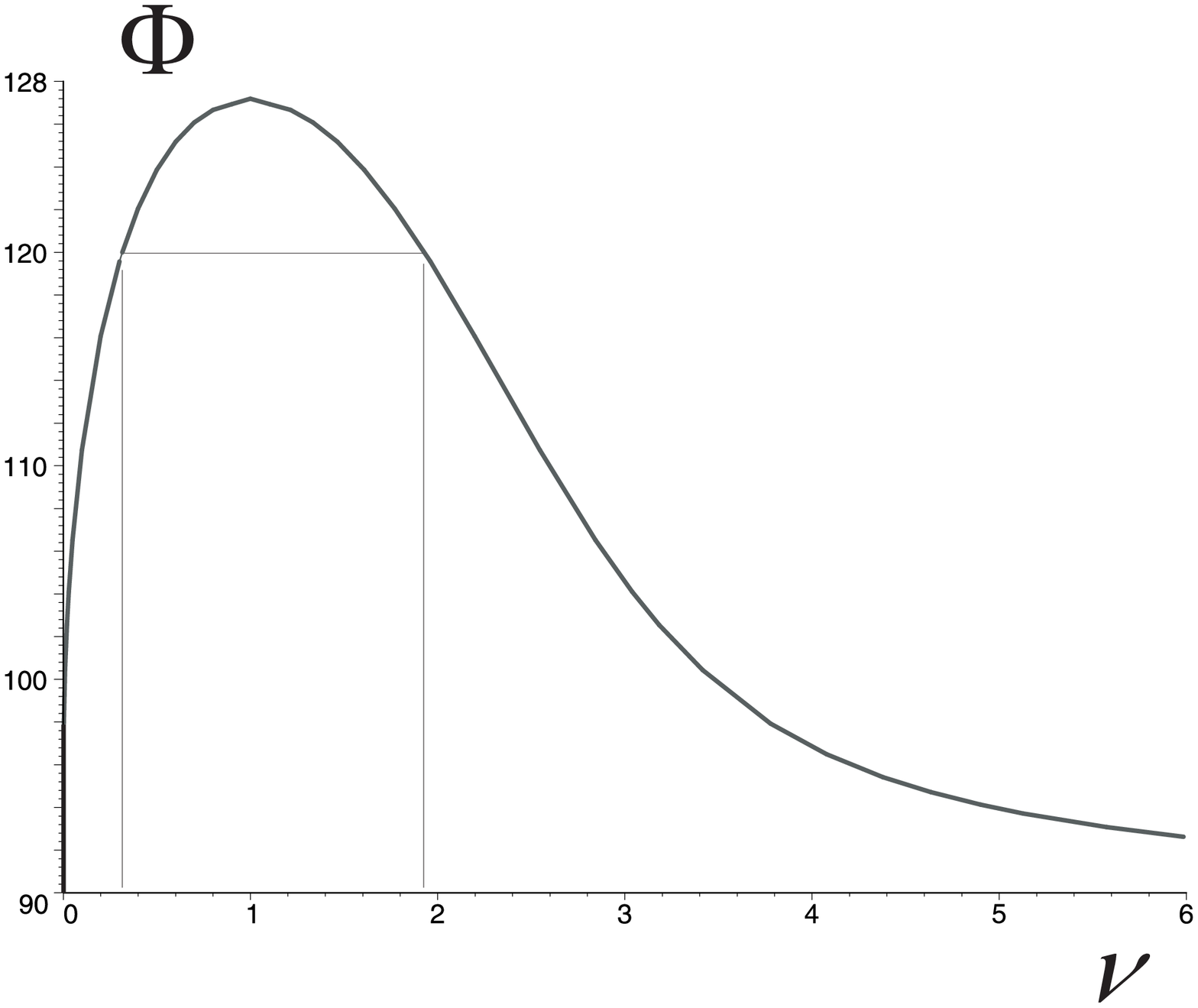,width=62mm,height=50mm}} 
\vspace*{3mm} 

\noindent {\small {\bf Bildo 5} {\it Angulo $\Phi$, en gradoj, inter sinsekvaj $\rho_{max}$ kaj  $\rho_{min}$, kiel funkcio de parametro $\nu$ de komenca pure azimuta velo. 
Se $\Phi=120^o$, ni havas amba\w $\nu\approx0,32$ kaj $\nu\approx1,92$, do bildo 4 montras trajektorion de movo kun $\nu\approx1,92$, anka\w. } } 
\vspace*{5mm}

En bildo 5 ni vidas ke $\nu\rightarrow0$ faras $\Phi\rightarrow90^o$. 
Tiu montras ke se parteto komencas movi de $\phi=0$ kun malgranda pure azimuta velo, 
sekvante \^gi iros preska\w rekte al altiranta akso $z$. 
\^Gi pasas \^cirka\w la akso kun  $\phi=\Phi\approx90^o$. 
Poste \^gi sekvas rekte \^gis la simetrika pozicio kun  $\phi\approx180^o$, 
kaj poste revenas al \^cirka\u{u}a\^\j o de komenca movo.  
 
Kiam $\nu\rightarrow1$ (kiam $v_0\rightarrow\sqrt{2G\lambda}$), okazas   $\Phi\rightarrow90^o\sqrt{2}\approx127^o$; tiu montras ke malgrandaj malordoj de ronda movo ka\u{u}zas oscilojn de parteto \^cirka\w la meza radio, kun ripetantaj interanguloj $2\Phi=180^o\sqrt{2}\approx254^o$.  
Fine, kiam $\nu\rightarrow\infty$ parteto iras de finita $\rho_{min}$ al  $\rho_{max}\rightarrow\infty$, atingata kun $\phi=\Phi\approx90^o$. 
Poste, parteto komencas movon de reveno, pasas \^cirka\w akso $z$ je distanco $\rho_{min}$ kun  $\phi\approx180^o$. Sekvante \^gi denove iras al $\rho_{max}\rightarrow\infty$, nun kun $\phi\approx270^o$, kaj poste revenas al \^cirka\u{u}a\^\j o de komenca pozicio.  

Bildo 6 montras grafikon de $\Phi$ kiel funkcio de racio $\rho_{max}/\rho_{min}$, pli precize de $\log_{10}(\rho_{max}/\rho_{min})$. 
  
\vspace*{3mm}
\centerline{\epsfig{file=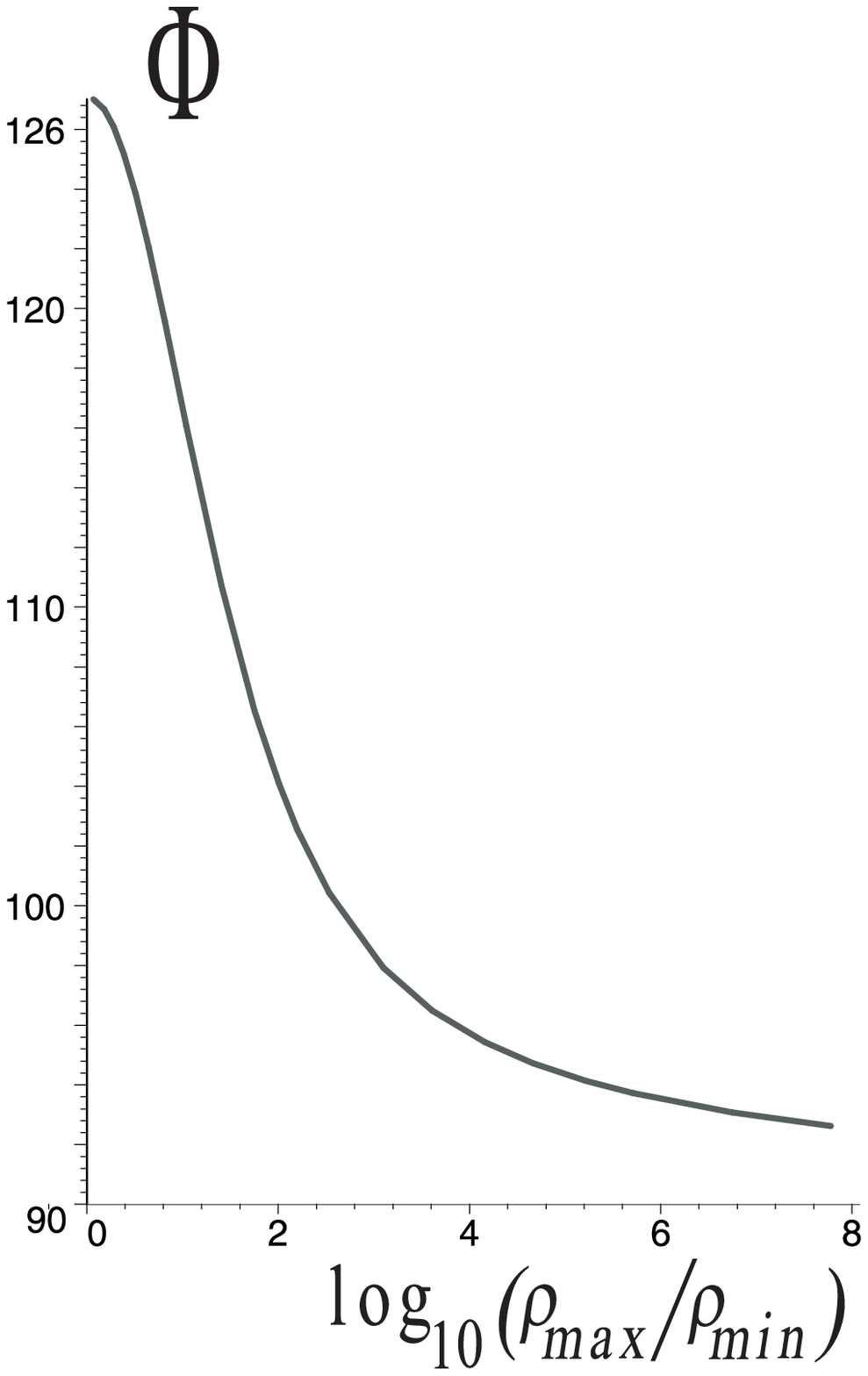,width=40mm,height=50mm}} 
\vspace*{3mm} 

\noindent {\small {\bf Bildo 6} {\it Angulo $\Phi$, en gradoj, inter sinsekvaj $\rho_{max}$ kaj  $\rho_{min}$, kiel funkcio de decimala logaritmo de racio $\rho_{max}/\rho_{min}$. La grafiko estas bona por amba\w $\nu<1$ kaj $\nu\geq1$. } } 
\vspace*{5mm}

Fermaj trajektorioj okazas se $p\Phi=q\pi$, kun $p$ kaj $q$ koprimaj entjeroj, kaj kun  $90^o<\Phi<90^o\sqrt{2}$.  
La plej elementaj havas  
\bea                                                                       \label{e17}
[p,q]=[3,2],\hskip1mm [5,3],\hskip1mm [7,4],\hskip1mm [8,5],\hskip1mm [9,5],\hskip1mm \dots 
\eea   
La trajektorio en bildo 4 havas la unuan duopon, $p$=3, $q$=2. 
Estus interesa entabeligi valorojn de $\nu$ kiuj okazas fermajn trajektoriojn, sed \^ci tie ni ne faros \^gin.  

Fine, kelkaj paroloj pri okazoj kun $\nu>1$ estas indaj. 
Tiuj okazoj havas komenca pure azimuta velo $v_0$ pli granda ol $v_c$, velo de rondaj movoj.  
Oni facile vidas ke movoj kun $\nu>1$ estas samaj al movoj kun $v_0<v_c$.  
Fakte, tiuj du movoj inter\^san\^gas rolon de $v_{max}$ kun $v_{min}$, kaj de $\rho_{max}$ kun $\rho_{min}$. 
Tiu jam evidenti\^gis per trajektorio kun $\Phi=120^o$, kiu okazas kun amba\w $\nu\approx0,32<1$ kaj $\nu\approx1,92>1$.   
\section{Movo en cilindra surfaco}
La plej \^generala movo de test-parteto estas kunmeto movanta en plano $z=$ konst (sekcio 5) kun unuforma movo paralela al akso $z$. 
Do, tiu movo okazas en rektata surfaco, paralela al akso $z$. 
 
La plej simpla tia surfaco estas ronda cilindra \^cirka\w akso $z$. 
Movo en tia cilindro estas ronda helico, kiel korktirilo, kun konstanta velo.  
La aliaj surfacoj havas sekciojn normalajn al akso $z$ kiel en sekcio 5 (plano $z=$ konst).
\section{Finaj rimarkoj}
En nia ser\^co, ankora\w iranta, pri geodezioj en kurba spaco-tempo de Levi-Civita, ni findis interesan kompari niajn rezultojn kun newtonaj movoj.  
Mirige, ni ne atingis findi en literaturo priskribon de tiuj movoj.  
Do ni decidis studi \^gin, kaj \^ci tie artikolo estas frukto de nia studo.   

Kelkaj rimarkoj estas eble utilaj, pri simetrioj de movoj sub la logaritma potencialo.  
Tri evidentaj simetrioj venas el la statikeco kaj la cilindreco de la gravita kampo.  
Resume ni dirus ke se kuna\^\j o  
\bea                                                                      \label{e18}    
\{\rho\!=\!f_\rho(t),\hskip2mm z\!=\!f_z(t),\hskip2mm \phi\!=\!f_\phi(t)\}
\eea 
reprezentas libera movo sub la kampo de $\lambda$, tiuokaze anka\w kuna\^\j o  
\bea                                                                      \label{e19}
\{\rho\!=\!f_\rho(t+k_t),\hskip2mm z\!=\!k_z+f_z(t+k_t),\hskip2mm \phi\!=k_\phi+\!f_\phi(t+k_t)\} 
\eea 
reprezentos, por iuj ajn valoroj de $k_t, k_z, k_\phi$. 

Qvara simetrio estis plurfoje jam sugestata, kaj nun estas klare elmetata:  
sub la kampo de $\lambda$, anka\w la kuna\^\j o  
\bea                                                                      \label{e20}
\{\rho\!=\!kf_\rho(k/t),\hskip2mm  z\!=\!kf_z(k/t),\hskip2mm  \phi\!=\!f_\phi(k/t)\} 
\eea 
reprezentos libera movo, por iu ajn valoro de $k$. 
Perparole, izotropika spacia ekspansio de iu ajn libera trajektorio $O$, kun centro de ekspansio sur la fonto, produktas novan liberan trajektorion $O'$. 
Estas klariginda kiel le\^go de velo bezonas okazi en $O'$, rilate al le\^go en $O$: 
la linia velo en punkto $P'\in O'$ bezonas esti sama ol en punkto $P\in O$, \^gia korespondanto per la ekspansio.  
Bildo 7 montras tiu simetrio, por okazo de pure radia movo.   

\vspace*{3mm}
\centerline{\epsfig{file=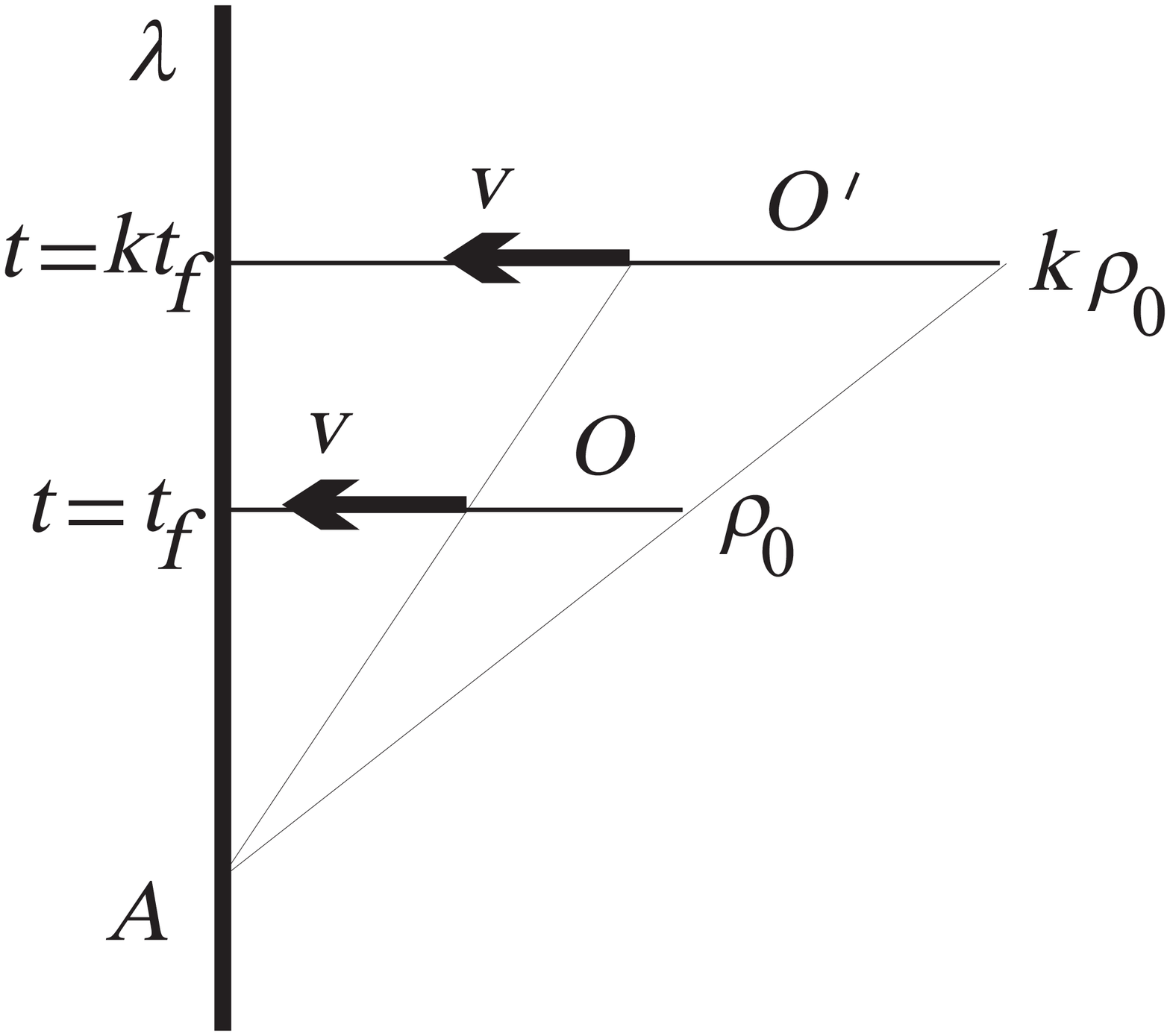,width=4cm,height=4cm}} 
\vspace*{3mm} 

\noindent {\small {\bf Bildo 7} {\it $O'$ estas pligrandiga trajektorio de pure radia trajektorio $O$, kun centro de ekspansio $A$ sur akso $z$.   
Amba\w $O'$ kaj $O$ korespondas a liberaj movoj. 
Velo de falo en $k\rho\in O'$ kaj en $\rho\in O$ estas samaj. }}  
\vspace*{5mm} 

Ni demandas se trajektorio $O$ de libera movo, kun sia le\^go de linia (tangencia) velo, povus havi alia le\^go de velo.  
Respondo esta jes, kondi\^ce ke anka\w la gravita fonto estu modifata.  
Se le\^go de velo $v(\rho,z,\phi)$ \^san\^gas al $kv(\rho,z,\phi)$, kun $k$=konst, tiuokaze la fonto $\lambda$ devas \^san\^gi al $k^2\lambda$. 
Anka\w en hiperbola potencialo (sfera simetrio), kaj en linia potencialo (plana simetrio), povas okazi sama trajektorio kun nesamaj liberaj movoj.  
Se \^san\^go estas $v(x,y,z)\rightarrow kv(x,y,z)$, bezonas \^san\^gi $m\rightarrow k^2m$ kaj $\sigma\rightarrow k^2\sigma$, respektive. 

Alia domando estus: \^cu anka\w en potencialoj $-Gm/r$ kaj $G\sigma h$ ekzistas 'kva\-ran si\-me\-tri\-on' (rilate izotropika ekspansio), kiel la (\ref{e20})? 
Denove respondo estas jes, kondi\^ce ke amba\w gravitaj fontoj kaj le\^goj de movo estu modifataj. 
 
En sfera simetrio, se $r\rightarrow kr$, tiuokaze devas \^san\^gi $v\rightarrow k^\alpha v$ kaj $m\rightarrow k^\beta m$ kun $2\alpha=\beta-1$. 
Kaj en plana simetrio, se ekspansio havas faktoro $k$, tiuokazo devas \^san\^gi $v\rightarrow k^\alpha v$ kaj $\sigma\rightarrow k^\beta\sigma$ kun $2\alpha=\beta+1$.
La centro $A$ de ekspansio estas sur maso $m$ en sfera okazo, kaj estas sur $\sigma$, en la plana.  

Programoj Maple, Corel, PCTeX, kaj eps.fig, estis uzataj por kunmeti \^ci tiu artikolo.   
Ni kore dankas al \^giaj kreintoj, same al ekipo de Microsoft kaj de Arxiv/LANL.  

\end{document}